\documentclass[
	conference,
]{IEEEtran}

\IEEEoverridecommandlockouts 
\makeatletter
\def\markboth#1#2{\def\leftmark{\@IEEEcompsoconly{\sffamily}\MakeUppercase{\protect#1}}%
\def\rightmark{\@IEEEcompsoconly{\sffamily}\MakeUppercase{\protect#2}}}
\makeatother

\usepackage{pbox}
\usepackage[amssymb]{SIunits}
\usepackage[cmex10]{amsmath}
\usepackage{amssymb}
\usepackage{amsthm}
\usepackage[latin1]{inputenc}
\usepackage{cite}
\usepackage{url}
\usepackage{epstopdf}
\usepackage{glossaries}
\usepackage{graphicx}
\usepackage[hang,small,bf]{caption}
\usepackage[subrefformat=parens,labelformat=parens]{subfig}
\usepackage{bm}
\usepackage[usenames]{color}
\usepackage[lined,ruled,linesnumbered]{algorithm2e}
\usepackage{balance}

\newenvironment{DIFnomarkup}{}{}

\newcommand{\eqsref}[2]{(\ref{#1})--(\ref{#2})}	
\newcommand{\figref}[1]{Fig.~\ref{#1}}

\newcommand{\tabref}[1]{Table~\ref{#1}}

\newcommand{\secref}[1]{Section~\ref{#1}}

\newtheorem{theorem}{Theorem}

\renewcommand{\Pr}[1]{\mathrm{Pr}\left\{#1\right\}}

\newcommand{\setU}{\mathcal{U}}

\newcommand{\PEPt}{\bar{p}}

\newcommand{\setV}{\mathcal{V}}
\newcommand{\setC}{\mathcal{C}}
\newcommand{\setE}{\mathcal{E}}
\newcommand{\setS}{\mathcal{S}}
\newcommand{\setG}{\mathcal{G}}
\newcommand{\setA}{\mathcal{A}}

\newcommand{\maxd}{q}

\newcommand{\uA}{\mathsf{A}}
\newcommand{\uB}{\mathsf{B}}

\newacronym{AWGN}{AWGN}{additive white Gaussian noise} 
\newacronym{TCP}{TCP}{transmission control protocol} 
\newacronym{CSA}{CSA}{coded slotted ALOHA} 
\newacronym{ABCSA}{B-CSA}{all-to-all broadcast coded slotted ALOHA}
\newacronym{SNR}{SNR}{signal-to-noise ratio} 
\newacronym{SINR}{SINR}{signal-to-interference-plus-noise ratio}
\newacronym{PLR}{PLR}{packet loss rate} 
\newacronym{UEP}{UEP}{unequal error protection} 
\newacronym{BS}{BS}{base station} 
\newacronym{LDPC}{LDPC}{low-density parity-check} 
\newacronym{VN}{VN}{variable node} 
\newacronym{CN}{CN}{check node} 
\newacronym{DE}{DE}{density evolution} 
\newacronym{MDS}{MDS}{maximum distance separable} 
\newacronym{BEC}{BEC}{binary erasure channel} 
\newacronym{PEC}{PEC}{packet erasure channel} 
\newacronym{DAMA}{DAMA}{demand assignment multiple access} 
\newacronym{CSMA}{CSMA-CA}{carrier sense multiple access with collision avoidance} 
\newacronym{VANET}{VANET}{vehicular ad hoc network} 
\newacronym{V2V}{V2V}{vehicle to vehicle} 
\newacronym{PHY}{PHY}{physical layer} 
\newacronym{MAC}{MAC}{medium access control} 
\newacronym{ARQ}{ARQ}{automatic repeat request} 
\newacronym{CDMA}{CDMA}{code division multiple access} 
\newacronym{TDMA}{TDMA}{time division multiple access}
\newacronym{CAM}{CAM}{cooperative awareness message}
\newacronym{DENM}{DENM}{decentralized environmental notification message}
\newacronym{ETSI}{ETSI}{European Telecommunications Standards Institute}
\newacronym{GPS}{GPS}{Global Positioning System}
\newacronym{DUEP}{DUEP}{double unequal error protection}
\newacronym{VC}{VC}{vehicular communication}
\newacronym{RU}{RU}{receiving user}
\newacronym{SIC}{SIC}{successive interference cancellation}

\graphicspath{%
	{../pstricks/}%
}%

\include{acronyms}
\include{hyphenations}
\include{newcommands}

\begin{document}

\begin{DIFnomarkup}

\title{Probabilistic Handshake in All-to-all Broadcast Coded Slotted ALOHA}

\author{%
\IEEEauthorblockN{Mikhail~Ivanov, Petar Popovski{\IEEEauthorrefmark{4}}, Fredrik Br\"{a}nnstr\"{o}m, Alexandre Graell i Amat, and \v Cedomir Stefanovi\' c{\IEEEauthorrefmark{4}}\\}
\IEEEauthorblockA{Department of Signals and  Systems, Chalmers  University of Technology, Gothenburg, Sweden\\} \IEEEauthorblockA{ \IEEEauthorrefmark{4}Department of Electronic Systems, Aalborg University, Aalborg, Denmark\\ 
\emph{\{mikhail.ivanov, fredrik.brannstrom, alexandre.graell\}@chalmers.se, \{petarp, cs\}@es.aau.dk}
}
\thanks{This research was supported by the Swedish Research Council, Sweden, under Grant No. 2011-5950, the Ericsson's Research Foundation, Sweden, Chalmers Antenna Systems Excellence Center in the project `Antenna Systems for V2X Communication', and the Danish Council for Independent Research, under Grants No. 11-105159 and No. 4005-00281. 
}

}%


\maketitle

\end{DIFnomarkup}

\begin{abstract}

We propose a probabilistic handshake mechanism for all-to-all broadcast coded slotted ALOHA. We consider a fully connected network where each user acts as both transmitter and receiver in a half-duplex mode. Users attempt to exchange messages with each other and to establish one-to-one handshakes, in the sense that each user decides whether its packet was successfully received by the other users: After performing decoding, each user estimates in which slots the resolved users transmitted their packets and, based on that, decides if these users successfully received its packet. The simulation results show that the proposed handshake algorithm allows the users to reliably perform the handshake. The paper also provides some analytical bounds on the performance of the proposed algorithm which are in good agreement with the simulation results.
\end{abstract}

\glsresetall

\section{Introduction}\label{sec:intro}

\Glspl{VC} is presently one of the most challenging problems of communication engineering. Its deployment will enable numerous applications, such as intelligent transportation systems, autonomous driving, and, most importantly, traffic safety. \glspl{VC} entails a number of challenges, such as all-to-all communication, high mobility networks with rapidly changing topologies and a large number of users, and poor channel quality. These challenges require new ideas and designs at the physical and the \gls{MAC} layers. The main requirements for \glspl{VC} are high reliability and low latency. Furthermore, the aforementioned challenges prohibit the classical use of acknowledgements in the form of additional signaling.

A novel \gls{MAC} protocol called \gls{ABCSA} was proposed by the authors in~\cite{Ivanov15}, which was shown to be able to satisfy the reliability and latency requirements in rough conditions of vehicular networks under a set of idealized assumptions, such as perfect interference cancellation. Originally proposed for a unicast scenario, \gls{CSA} can provide large throughputs close to those of coordinated schemes~\cite{Paolini11,Stefanovic13}. Different versions of \gls{CSA} have been proposed (see~\cite{Paolini15} for the most recent review). All of them share a slotted structure borrowed from the original slotted ALOHA~\cite{Roberts75} and the use of successive interference cancellation. The contending users introduce redundancy by encoding their messages into multiple packets, which are transmitted in randomly chosen slots. In the unicast scenario, the \gls{BS} buffers the received signal, decodes the packets from the slots with no collision and attempts to reconstruct the packets in collision exploiting the introduced redundancy. A packet that is reconstructed is subtracted from the buffered signal and the \gls{BS} proceeds with another decoding round.

In contrast to classical \gls{CSA}, where a \gls{BS} is the intended recipient of the messages,  in~\gls{ABCSA} each user acts as both receiver and transmitter. Every user is equipped with a half-duplex transceiver, so that a user cannot receive packets in the slots it uses for transmission. This can be modeled as a packet erasure channel~\cite{Ivanov14} and it affects the design and the performance analysis as compared to classical \gls{CSA}.

Whereas \gls{ABCSA} can provide high reliability, the rear communication failure events may be extremely costly in safety applications. Since providing error-free communication under the described conditions of \glspl{VC} is not possible, one may attempt to detect communication failure events to use this information in the application level. For instance, consider the scenario where two users $\uA$ and $\uB$ are heading towards each other. If $\uA$ receives a message from $\uB$ and obtains the information that $\uB$ failed to receive its message, $\uA$ can take extra precautions to avoid collision with $\uB$. In this paper, we propose an algorithm to obtain this information based on the by-product of decoding, i.e., no extra signaling is used by the users. In particular, $\uA$ uses the knowledge of $\uB$'s transmissions to detect that $\uB$ failed to resolve $\uA$.

The problem studied in this paper resembles the one of handshake used for establishing connections in connection-oriented protocols. In \gls{TCP}, a three-way handshake is used by a pair of users to exchange their messages and to acknowledge that the messages were received~\cite{Sunshine78}. If the messages in the TCP level are exchanged, then the handshake is always performed successfully. In the proposed algorithm, however, the decision about successful handshake can be in error, which indicates its probabilistic nature.


\section{System Model}\label{sec:syst_model}
In this section, we first describe how encoding and decoding are performed in \gls{ABCSA}. Based on that, we describe the proposed handshake algorithm.
\subsection{Coded Slotted ALOHA}

\begin{figure}
	\centering
	\includegraphics{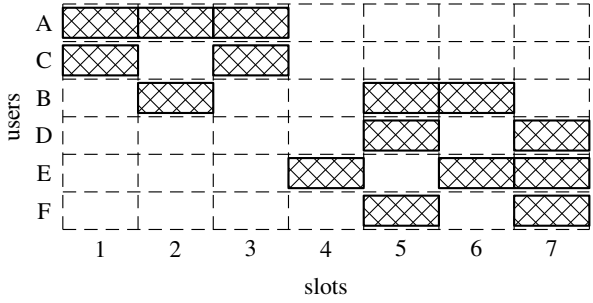}
	\caption{Users' transmissions in a B-CSA system within one frame. Shaded rectangles represent transmitted packets.}
	\label{fig:system_model}
\end{figure}

We consider a fully connected network with $m$ users that want to communicate between each other over a shared medium. We focus on the exchange of~\glspl{CAM}~\cite{etsi_cam} used for safety applications, which are transmitted periodically by each user. The transmission period is called \emph{frame} and it is divided into $n$ slots of equal duration. Users are assumed to be frame-synchronized by means of \gls{GPS}. Each user maps its message to a physical layer packet and repeats it $l$ times ($l$ is a random number chosen based on a predefined distribution) in randomly chosen slots, as shown in~\figref{fig:system_model} for a system with 6 users and 7 slots. Such a user is called a degree-$l$ user. Every packet contains pointers to its copies, so that, once a packet is successfully decoded, full information about the location of the copies is available.

\begin{figure}
	\centering
	
	\subfloat[Original ``unicast'' graph $\setG$.]{\includegraphics{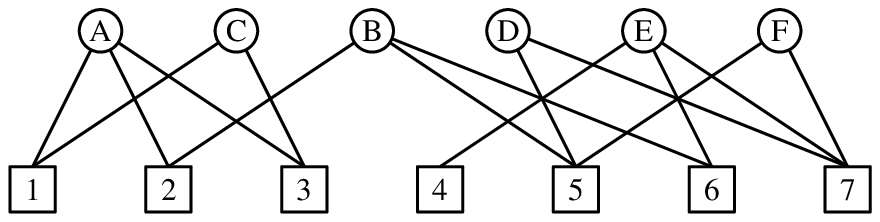}\label{subfig:graph1}}
	
	\subfloat[Induced graph $\setG_\uA$ for user $\uA$. $\setG_\uA(\uB) = 1$.]{\includegraphics{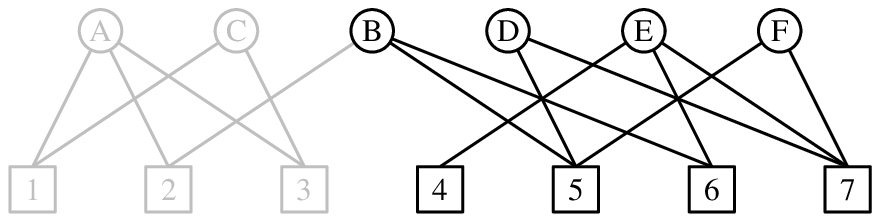}\label{subfig:graph2}}
	
	\subfloat[Intermediate graph $\setA''$ reconstructed by user $\uA$ and slots with residual interference (marked with gray).]{\includegraphics{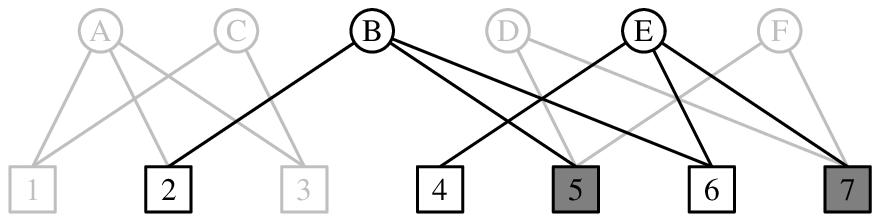}\label{subfig:graph3}}
	
	\subfloat[Intermediate graph $\setA'$ reconstructed by user $\uA$ excluding slots with residual interference.]{\includegraphics{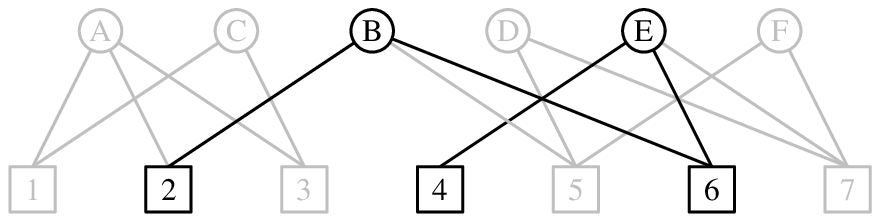}\label{subfig:graph4}}
	
	\subfloat[Graph $\setA$ reconstructed by user $\uA$ (including user $\uA$'s slots).]{\includegraphics{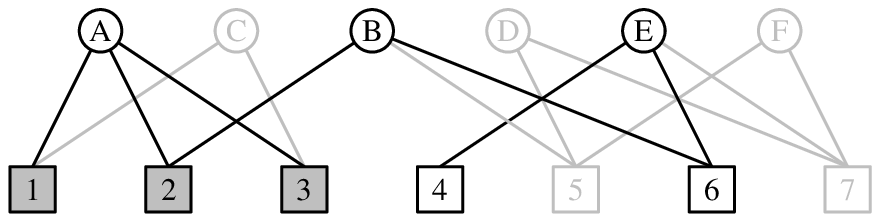}\label{subfig:graph5}}
	
	\subfloat[Induced graph $\setA_\uB$ for user $\uB$ based on $\setA$. $\setA_\uB(\uA) = 1.$]{\includegraphics{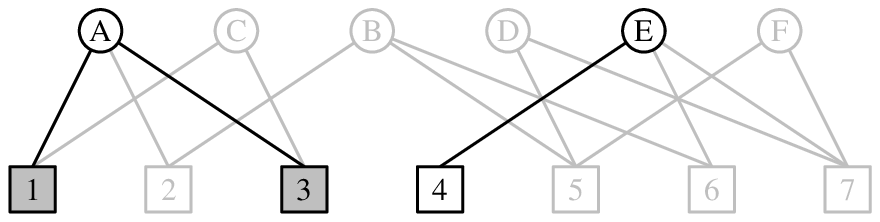}\label{subfig:graph6}}
	
		\subfloat[Induced graph $\setG_\uB$ for user $\uB$. $\setG_\uB(\uA) = 0.$]{\includegraphics{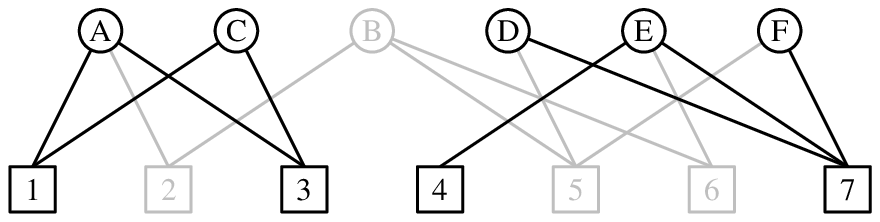}\label{subfig:graph7}}
	
	\caption{The graph evolution over the course of the handshake algorithm. Circles represent users (VNs) and squares represent slots (CNs). The example corresponds to the outcome $\boldsymbol{g} = [1,\,1,\,0]$ (see \secref{sec:outputs}).}
	\label{fig:graph}
\end{figure}

Under the assumptions described in the following, the system can be analyzed using the theory of codes on graphs on the \gls{BEC}. Each user corresponds to a \gls{VN} in the bipartite graph and represents a repetition code, whereas slots  correspond to \glspl{CN} and can be seen as single parity-check codes. In the following, users and \glspl{VN} are used interchangeably. An edge connects the $j$th \gls{VN} to the $i$th \gls{CN} if the $j$th user transmits in the $i$th slot. For the example in~\figref{fig:system_model}, the corresponding bipartite graph is shown in~\figref{fig:graph}\subref{subfig:graph1}. A bipartite graph is defined as $\setG =\{\setV, \setC, \setE\}$, where $\setV$, $\setC$, and $\setE$ represent the sets of \glspl{VN}, \glspl{CN}, and edges, respectively. 

The performance of the system depends on the distribution that users use to choose the degree $l$ or, using graph terminology, on the \gls{VN} degree distribution
\begin{equation}\label{eq:distr_orig}
	\lambda(x) = \sum_{l = 0}^{\maxd}\lambda_{l}x^{l},
\end{equation}
where $x$ is a dummy variable, $\lambda_l$ is the probability of choosing degree $l$, and $\maxd$ is the maximum degree, which is often bounded due to implementation constraints.  

Users buffer the received signal whenever they are not transmitting. The difference between classical \gls{CSA} and \gls{ABCSA} is illustrated in the example of~\figref{fig:graph}. The entire graph $\setG$ is available to the \gls{BS} in \gls{CSA} (\figref{fig:graph}\subref{subfig:graph1}). For a generic user $\uA$ in \gls{ABCSA}, the part of the graph that corresponds to user $\uA$'s transmissions is not available to it due to the half-duplex mode. Thus, this part of the graph shown with gray in~\figref{fig:graph}\subref{subfig:graph2}. The available part of the graph is called the \emph{induced} graph and is denoted by $\setG_{\uA}$
. The received signal buffered by user $\uA$ in slot $i$ is 
\begin{equation}
	y_i = \sum_{j \in \setU_i} h_{i,j} a_{j},
\end{equation}
where $\setU_i \subset \mathcal{U}$ is the set of users that transmit in the $i$th slot, $\mathcal{U}$ is the set of all users, $a_j$ is a packet of the $j$th user in $\mathcal{U}_i$, and $h_{i,j} > 0$ is the channel coefficient. A slot is called a \emph{singleton} slot if it contains only one packet. If it contains more packets, we say that a collision occurs.

When decoding, user $\uA$ first decodes the packets in singleton slots and obtains the location of their copies. Using data-aided methods, the channel coefficients corresponding to the copies are then estimated. After subtracting the interference caused by the identified copies, decoding proceeds until no further singleton slots are found.

The performance parameters of \gls{ABCSA} are defined as follows. The channel load is defined as $g = m/n$. The average number of users that are not successfully resolved by user $\uA$, termed \emph{unresolved users}, is denoted by $\bar{w}$. The reliability is measured by means of the average \gls{PLR}, $\PEPt = \bar{w}/(m-1)$, which is the probability of a user to be unresolved by user $\uA$.

\subsection{Probabilistic Handshake}
%
One of the main differences of \gls{CSA} compared to actual codes on graphs is that, in \gls{CSA}, the graph is not known to the decoder a priori and the decoder reconstructs it while decoding. We use this reconstructed graph to perform the handshake. Without loss of generality, we concentrate on the handshake between users $\uA$ and $\uB$.

We first introduce the necessary notation to describe the proposed handshake algorithm. Given a particular realization of the graph $\setG$ with $m$ \glspl{VN} and $n$ \glspl{CN} generated randomly using the distribution $\lambda(x)$, the reconstruction of the graph $\setG$ obtained by user $\uA$ after decoding is denoted by $\setA$. We recall that the induced graph for user $\uA$ is denoted by $\setG_{\uA}$. With a slight abuse of notation, if user $\uB$ is resolvable by user $\uA$ based on $\setG_{\uA}$, we write $\setG_{\uA}(\uB) = 1$ and we write $\setG_{\uA}(\uB) = 0$ otherwise. Using this notation, the \gls{PLR} can be written as $\bar{p} = \Pr{\setG_{\uA}(\uB) = 0}$. $\setA_{\uB}$ denotes the graph that user $\uA$ obtains after removing user $\uB$'s slots from the reconstructed graph $\setA$. Hence, we write $\setA_{\uB}(\uA) = 1$ if user $\uA$ concludes that user $\uB$ resolves it using the reconstructed graph $\setA$.

For the example in~\figref{fig:graph}, user $\uA$ uses the induced graph $\setG_{\uA}$ shown in~\figref{fig:graph}\subref{subfig:graph2} for decoding. However, user $\uA$ may not be able to fully reconstruct $\setG$. In fact, user $\uA$ can only resolve users $\uB$ and $\mathsf{E}$. Users $\mathsf{D}$ and $\mathsf{F}$ cannot be resolved because they form a so-called \emph{stopping set}, a harmful graph structure that makes decoding fail. A stopping set is a subset of  \glspl{VN} of non-zero degrees $\setS \subset \setV$ where all neighboring \glspl{CN} of $\setS$ are connected to $\setS$ at least twice~\cite{Di02}. 

After decoding, user $\uA$ reconstructs the graph $\setA''$ shown in~\figref{fig:graph}\subref{subfig:graph3}. Additionally, user $\uA$ obtains the knowledge that slots 5 and 7 belong to a stopping set. It is worth noting that this is a stopping set from user $\uA$'s perspective and not necessarily a stopping set in the original graph $\setG$. Nonetheless, user $\uA$ assumes that these slots cannot be used for decoding by any other user. Therefore, user $\uA$ removes these slots, as well as all the  edges connected to them, to obtain the graph $\setA'$ shown in~\figref{fig:graph}\subref{subfig:graph4} (the removed slots and edges are shaded).

As the the last step to reconstruct $\setG$, user $\uA$ adds itself as a VN to the graph $\setA'$ and connects it to the corresponding CNs. The reconstruction of the graph $\setA$ is shown in~\figref{fig:graph}\subref{subfig:graph5}. Since user $\uA$ does not know who exactly transmitted in slots 1--3, these slots are shown with gray. User $\uA$ uses this graph to run the decoding on behalf of other users, e.g., \figref{fig:graph}\subref{subfig:graph6} shows the graph $\setA_{\uB}$ that user $\uA$ uses for decoding on behalf of user $\uB$. In this case, $\setA_{\uB}(\uA) = 1$. However, in reality user $\uB$ uses $\setG_{\uB}$ for decoding and its true outcome is $\setG_{\uB}(\uA) = 0$. Therefore, user $\uA$ makes an erroneous decision about user $\uB$'s awareness of user $\uA$, which happens due to partial knowledge about the slots user $\uA$ uses for transmission.

\subsection{Handshake Outcomes}\label{sec:outputs}
If user $\uA$ is in a stopping set contained in the original graph $\setG$, then it will not be resolvable by any other user in the network. User $\uA$ has no means to learn about this since this information is contained in the slots that it uses for transmission. Hence, user $\uA$ can never be sure about its successful handshake decision. To describe the possible outcomes of the handshake algorithm and analyze their probabilities, we introduce the vector $\boldsymbol{g} = [\setG_\uA(\uB), \,\setA_\uB(\uA),\,\setG_\uB(\uA)]$. If user $\uA$ fails to resolve user $\uB$, i.e., $\setG_\uA(\uB) = 0$, with a slight abuse of notation we say that $\setA_\uB(\uA) = \mathsf{x}$, meaning that $\uA$ cannot perform a handshake with $\uB$.  All possible outcomes with the corresponding probabilities are summarized in~\tabref{tab:outcomes}. In the table, $p_1$ is the probability that user $\uA$ successfully detects communication failure at user $\uB$'s side. $p_2$ is the probability that user $\uA$ fails to detect communication failure at user $\uB$'s side and erroneously assumes that user $\uB$ successfully receives its packet. $p_5$ is the probability of correct handshake. $p_3$ and $p_4$ are auxiliary probabilities, where $p_3 = \Pr{\setG_\uB(\uA) = 0 , \setG_\uA(\uB) = 0}$ is the probability that users $\uA$ and $\uB$ do resolve each other simultaneously. The sum $p_3 +p_4$ equals the probability that user $\uA$ does not resolve user $\uB$, i.e., $p_3 + p_4 = \Pr{\setG_\uA(\uB) = 0} = \bar{p}$.

\begin{table}
\caption{Possible outcomes of the handshake algorithm.\\ $\boldsymbol{g} = [\setG_\uA(\uB), \,\setA_\uB(\uA),\,\setG_\uB(\uA)]$.}
\begin{center}
  \begin{tabular}{c|c|c|c|c}
	\hline
	$\setG_\uA(\uB)$ & $\setA_\uB(\uA)$ & $\setG_\uB(\uA)$&$\Pr{\boldsymbol{g}}$& Event\\
	\hline
	$1$ & $0$ & $0$ &$p_1$& Failure detected\\
	\hline
	$1$ & $1$ & $0$ &$p_2$& False handshake \\
	\hline
	$0$ & $\mathsf{x}$ & $0$ &$p_3$& Auxiliary\\
	\hline
	$0$ & $\mathsf{x}$ & $1$ &$p_4$& Auxiliary\\
	\hline	    
	$1$ & $1$ & $1$&$p_5$& Successful handshake\\
	\hline
  \end{tabular}
  \end{center}
  \label{tab:outcomes}
\end{table}

Interestingly, the outcome $\boldsymbol{g} = [1,\,0,\, 1]$ can not occur, which is formally proven in the following theorem.

\begin{theorem}\label{theor:impossible_ss}
$\Pr{\boldsymbol{g} = [1,\,0,\, 1]} = 0$.
\end{theorem}
\begin{IEEEproof}
	$\setG_\uA(\uB) = 1$ and $\setA_\uB(\uA) = 0$ imply that user $\uA$ is in a stopping set $\setS$ of the graph $\setA_\uB$. This stopping set $\setS$ has to be present in the graph $\setG_\uB$ as well since $\setA_\uB$ is a subgraph of $\setG_\uB$. Hence, $\setG_\uB(\uA) = 0$, which completes the proof.
\end{IEEEproof}

\section{Performance Analysis}\label{sec:analysis}

\subsection{Induced Distribution}\label{sec:ind_distr}
The performance of \gls{CSA} exhibits a threshold behavior, i.e., all users are successfully resolved if the channel load is below a certain threshold value when $n \rightarrow \infty$. The threshold depends only on the degree distribution and is obtained via density evolution~\cite{Liva11}. A finite number of slots, however, gives rise to an error floor in the \gls{PLR} performance. In~\cite{Ivanov15} it was shown that the error floor can be accurately predicted based on the induced distribution observed by the receiver. The induced distribution for a degree-$k$ receiver can be expressed similarly to~\eqref{eq:distr_orig}, where
\begin{equation}
	\lambda^{(k)}_d = \sum_{l = d}^{\min\{q, k+d\}}   \frac{\binom{n-k}{d}\binom{k}{l-d}}{\binom{n}{l}}\lambda_{l}\label{eq:t_induced}
\end{equation}
is the fraction of users of degree $d$ as observed by user $\uA$ if it chooses degree $k$.


We define the \gls{PLR} for a degree-$d$ user as observed by a degree-$k$ receiver as
\begin{equation}\label{eq:pep_l}
\bar{p}^{(k)}_d = \frac{\bar{w}^{(k)}_d}{\bar{m}^{(k)}_d} = \frac{\bar{w}^{(k)}_d}{ m \lambda^{(k)}_d},
\end{equation}
where $\bar{m}^{(k)}_d$ and $\bar{w}^{(k)}_d$ are the average number of all and unresolved degree-$d$ users for a degree-$k$ receiver, respectively. For degree-$0$ users, $p^{(k)}_0=1$ for all $k$.

\subsection{Analytical Results}
In this section, we are interested in describing the probabilities $p_1$ and $p_2$. First, we observe that 
\begin{equation}\label{eq:psum}
	p_1 + p_2 + p_3 = \Pr{\setG_{\uB}(\uA) = 0} =  \bar{p}.
\end{equation}
From~\eqref{eq:psum}, we can immediately write that
\begin{equation}\label{eq:bound1}
	p_1 + p_2 \le  \bar{p}.
\end{equation}
In the following, we tighten the bound in~\eqref{eq:bound1}. The probability $p_3$ can be written as
\begin{multline}\label{eq:p3}
p_3 = \Pr{\setG_\uA(\uB) = 0} \Pr{\setG_\uB(\uA) = 0 | \setG_\uA(\uB) = 0}.
\end{multline}
The fact that the induced graphs for all users arise from the same original graph $\setG$ gives rise to dependency between users' performance. This is expressed as the conditional probability in the right-hand side of~\eqref{eq:p3}.

Examining~\eqref{eq:p3}, we conjecture that 
\begin{equation}\label{eq:conj}
p_3 \ge \bar{p}^2.
\end{equation}
For the asymptotic case, when $n \rightarrow \infty$, it is easy to show that $p_3 = \bar{p}^2$ since the probability for users $\uA$ and $\uB$ to use the same slots for transmission is zero. 

For finite frame lengths, the rationale behind this conjecture is as follows. Let us take a closer look at the conditional probability in~\eqref{eq:p3}. We conjecture that
\begin{equation} \label{eq:condition_aux}
\Pr{\setG_\uB(\uA) = 0 | \setG_\uA(\uB) = 0} \ge \Pr{\setG_\uB(\uA) = 0}.
\end{equation}
Using the law of total probability, we can write
\begin{align*}
&\Pr{\setG_\uB(\uA) = 0 | \setG_\uA(\uB) = 0} \Pr{\setG_\uA(\uB) = 0}\\
 +& \Pr{\setG_\uB(\uA) = 0 | \setG_\uA(\uB) = 1} \Pr{\setG_\uA(\uB) = 1}\\
  = & \Pr{\setG_\uB(\uA) = 0}.
\end{align*}
Exploiting $\Pr{\setG_\uB(\uA) = 0} = \Pr{\setG_\uA(\uB) = 0} = \bar{p}$ gives
\begin{multline}
\Pr{\setG_\uB(\uA) = 0 | \setG_\uA(\uB) = 1} \\= \frac{\bar{p}}{1 - \bar{p}} \left( 1 - \Pr{\setG_\uB(\uA) = 0 | \setG_\uA(\uB) = 0} \right).
\end{multline}
Therefore, showing~\eqref{eq:condition_aux} is equivalent to showing
\begin{equation}\label{eq:condition}
\Pr{\setG_\uB(\uA) = 0 | \setG_\uA(\uB) = 1} \le \Pr{\setG_\uB(\uA) = 0}.
\end{equation}

Assuming an \gls{UEP} property~\cite{Ivanov14}, i.e., $\bar{p}^{(k)}_{l+1} < \bar{p}^{(k)}_l$ for a given $k$, it can be shown that~\eqref{eq:condition} and, hence,~\eqsref{eq:conj}{eq:condition_aux} hold. The derivations are omitted due to lack of space. The \gls{UEP} property does hold when $n \rightarrow \infty$. It is also easy to find a counterexample for extremely short frame lengths, when it does not hold. For instance, consider a unicast system ($k = 0$) with two users, two slots, and the distribution $\lambda(x) = 0.5x + 0.5x^2$. For this toy example, the \gls{PLR} for users of different degrees can be found by hand, yielding $\bar{p}^{(0)}_1 = 0.25$ and $\bar{p}^{(0)}_2 = 0.5$. However, from our extensive simulations, we conjecture that the \gls{UEP} property does hold for sufficiently large values of $n$. Proving it rigorously and characterizing sufficient frame lengths is subject of ongoing work.

Using the conjecture in~\eqref{eq:conj} together with~\eqref{eq:psum}, we can write a tighter version of~\eqref{eq:bound1} as
\begin{equation}\label{eq:bound}
p_1 + p_2  \le \bar{p}(1 - \bar{p}).
\end{equation}
Next section presents numerical results and confirms the conjectures made in this section.



\section{Numerical Results}

\begin{figure}
	\centering 
	\subfloat[$\lambda(x) = 0.25 x^2 + 0.6 x^3 + 0.15 x^8$.]{
	\includegraphics[width=\columnwidth]{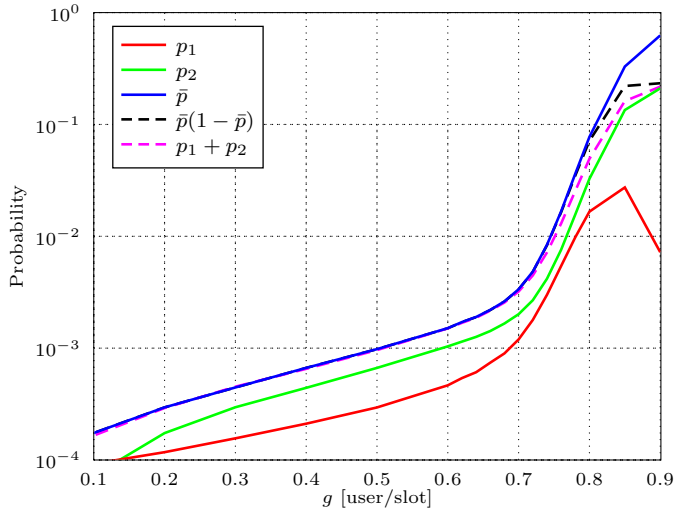}}

	\subfloat[$\lambda(x) = 0.86 x^3 + 0.14 x^8$.]{
	\includegraphics[width=\columnwidth]{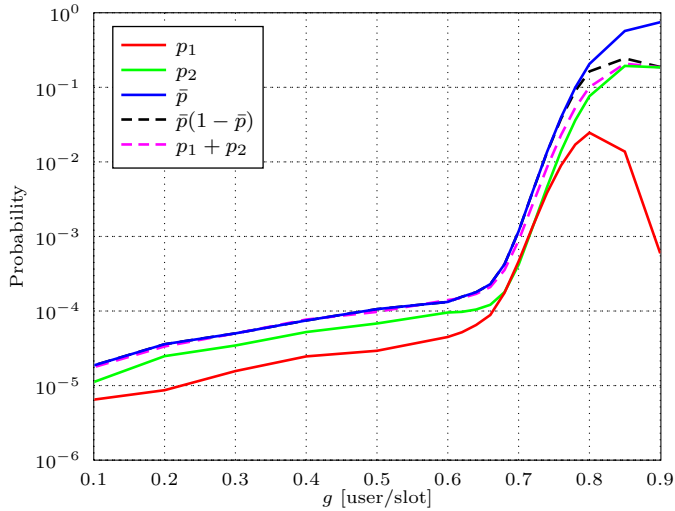}}
	\caption{Handshake performance of user $\uA$ in B-CSA for two different distributions and the frame length of  $n = 200$ slots.}
	\label{fig:handshake1}
\end{figure}

In~\figref{fig:handshake1}, we show simulation results for two different distributions and frame length $n =200$. The distribution $\lambda(x) = 0.25 x^2 + 0.6 x^3 + 0.15 x^8$ is taken from~\cite{Liva11}, where it was optimized for classical \gls{CSA} based on the threshold obtained via density evolution. The fraction of degree-2 users was limited to $0.25$ to yield low error floor.  The second distribution, $\lambda(x) = 0.86 x^3 + 0.14 x^8$, was optimized  in~\cite{Ivanov15} based on error floor approximations to provide low error floor for \gls{ABCSA}.

The red and the green curves show the probabilities $p_1$ and $p_2$, respectively, and characterize the handshake performance.  The blue curves show the \gls{PLR}. From~\figref{fig:handshake1} we observe that for low to moderate channel load
\begin{equation}\label{eq:sumpp_approx}
	p_1 + p_2  \approx \bar{p} (1- \bar{p}).
\end{equation}
Using~\eqref{eq:psum}, we conclude that $p_3 \approx \bar{p}^2$, which, using~\eqref{eq:p3}, leads to
\begin{equation}\label{eq:conditionl_ineq}
	\Pr{\setG_\uB(\uA) = 0 | \setG_\uA(\uB) = 0} \approx \bar{p}.
\end{equation}
Therefore, for low to moderate channel load, the probability for users to overlap in some slots is very small, which explains that the correlation between users' performance is negligible.

Bearing in mind that $\bar{p} \ll 1$ in the error floor region, we
\newpage
\noindent can further simplify~\eqref{eq:sumpp_approx} and write 
\begin{equation}\label{eq:sumpp_approx2}
	p_1 + p_2  \approx \bar{p}.
\end{equation}
In other words, the probability $\Pr{\setG_{\uB}(\uA) = 0} = \bar{p}$ consists of $p_1$ and $p_2$, so that user $\uA$ manages to detect communication failure events $\setG_{\uB}(\uA) = 0$ in $ p_1/\bar{p} \approx 30\%$ of the cases for both distributions. We remark, however, that this ratio may change depending on the distribution. This fact suggests a new design criterion for optimizing the degree distribution, i.e., the minimization of the probability of false handshake $p_2$. We also remark that the sum of $p_1 + p_2$, shown with dashed purple curves, is strictly smaller than $\bar{p} (1- \bar{p})$ (black dashed curves), which is in agreement with the bound in~\eqref{eq:bound}.

\section{Conclusions and Future Work}

In this paper, we proposed a probabilistic handshake algorithm for vehicular communications based on \gls{ABCSA}. In the rare cases of communication failure between two users, this event can be detected by one of the users. The simulation results show that around $30\%$ of such events can be detected for the considered distributions. We also proposed analytical bounds on the performance of the handshake algorithm, which match well the simulation results. The analytical bounds rely on the \gls{UEP} property of \gls{CSA}, for which a rigorous proof for finite frame lengths is left for future work.

\bibliographystyle{IEEEtran}

\end{document}